\documentclass[12pt,preprint]{aastex}
\usepackage{multirow}

\shorttitle{CO (2-1) in z$\sim$6 quasars}
\shortauthors{Wang et al.}
\slugcomment{\small Accepted for publication in the EVLA ApJL Special Issue.}
\begin{document}

\title{CO (2-1) Line Emission in Redshift 6 Quasar Host Galaxies}

\author{Ran Wang\altaffilmark{1,2,14},
Jeff Wagg\altaffilmark{1,3},
Chris L. Carilli\altaffilmark{1},
Fabian Walter\altaffilmark{4},
Dominik. A. Riechers\altaffilmark{5,15},
Chris Willott\altaffilmark{6},
Frank Bertoldi\altaffilmark{7},
Alain Omont\altaffilmark{8},
Alexandre Beelen\altaffilmark{9},
Pierre Cox\altaffilmark{10},
Michael A. Strauss\altaffilmark{11},
Jacqueline Bergeron\altaffilmark{8},
Thierry Forveille\altaffilmark{12},
Karl M. Menten\altaffilmark{13},
Xiaohui Fan\altaffilmark{2}}
\altaffiltext{1}{National Radio Astronomy Observatory, PO Box 0, Socorro, NM, 87801, USA}
\altaffiltext{2}{Steward Observatory, University of Arizona, 933 N Cherry Ave., Tucson, AZ, 85721, USA}
\altaffiltext{3}{European Southern Observatory, Alonso de C\'ordova 3107, Vitacura, Casilla 19001, Santiago 19, Chile}
\altaffiltext{4}{Max-Planck-Institute for Astronomy, K$\rm \ddot o$nigsstuhl 17, 69117 Heidelberg, Germany}
\altaffiltext{5}{California Institute of Technology, 1200 E. California Blvd., Pasadena, CA, 91125, USA}
\altaffiltext{6}{Herzberg Institute of Astrophysics, National Research Council, 5071 West Saanich Rd, Victoria, 
BC V9E 2E7, Canada}
\altaffiltext{7}{Argelander-Institut f$\rm \ddot u$r Astronomie, University of Bonn, Auf dem H$\rm \ddot u$gel 71, 53121 Bonn, Germany}
\altaffiltext{8}{Institut d'Astrophysique de Paris, CNRS and Universite Pierre et Marie Curie, Paris, France}
\altaffiltext{9}{Institut d'Astrophysique Spatiale, Universit$\rm\acute{e}$ Paris-Sud, F-91405 Orsay Cedex, France}
\altaffiltext{10}{Institute de Radioastronomie Millimetrique, St. Martin d'Heres, F-38406, France}
\altaffiltext{11}{Department of Astrophysical Sciences, Princeton University, Princeton, NJ, 08544, USA}
\altaffiltext{12}{Laboratoire d'Astrophysique, Observatoire de Grenoble, Universit$\rm\acute{e}$ J. Fourier, BP 53, F-38041 Grenoble, Cedex 9, France}
\altaffiltext{13}{Max-Planck-Institut f$\rm \ddot u$r Radioastronomie, Auf dem H$\rm \ddot u$gel 69, 53121 Bonn, Germany}
\altaffiltext{14}{Jansky Fellow}
\altaffiltext{15}{Hubble Fellow}
\begin{abstract}
We report new observations of CO (2-1) line emission
toward five z$\sim$6 quasars using the Ka-band receiver system on the Expanded Very Large Array (EVLA).
Strong detections were obtained in two of them, SDSS J092721.82+200123.7 and 
CFHQS J142952.17+544717.6, and a marginal detection was obtained in 
another source, SDSS J084035.09+562419.9. 
Upper limits of the CO (2-1) line emission have been obtained for the other two objects. 
The CO (2-1) line detection in J0927+2001, together with
previous measurements of the CO (6-5) and (5-4) lines, 
reveals important constraints on the CO excitation in the 
central $\sim$10 kpc region of the quasar host galaxy.
The CO (2-1) line emission from J1429+5447 is resolved into two distinct 
peaks separated by 1.2$''$ ($\rm \sim6.9$ kpc), indicating a possible 
gas-rich, major merging system, and the optical quasar position is consistent 
with the west peak. This result is in good agreement with the picture in which 
intense host galaxy star formation is coeval with rapid
supermassive black hole accretion in the most distant universe.
The two EVLA detections are ideal targets for further
high-resolution imaging (e.g., with ALMA or EVLA observations) to study the
gas distribution, dynamics, and SMBH-bulge mass relation in these
earliest quasar-host galaxy systems.
\end{abstract}
\keywords{galaxies: evolution --- galaxies: high-redshift --- molecular data 
--- galaxies: active --- radio lines: galaxies}

\section{Introduction}

Observations of high-redshift quasars probe the growth of supermassive
black holes (SMBH) and their connection to galaxy formation at the earliest
cosmic epochs. 
The discovery of strong submillimeter/millimeter [(sub)mm] dust 
continuum in about 30\% of the quasars known at z$\sim$6 provides the 
first evidence of active star formation in young quasar host 
galaxies at the end of the reionization era \citep{bertoldi03a,bertoldi03b,petric03,priddey03,
robson04,wang07,wang08}. The star formation rates 
estimated from the FIR luminosities (a few $\rm 10^{12}$ 
to $\rm 10^{13}\, L_{\odot}$) are on the order of $\rm 10^{2}$ 
to $\rm 10^{3}\,M_{\odot}\,yr^{-1}$, which are comparable to the typical 
values found in so-called submillimeter galaxies at $\rm z=2\sim3$ 
\citep{scott02,greve05,kovacs06}. The spatially resolved [C {\small II}] 
line emission from one of the most FIR luminous z$\sim$6 quasars, SDSS 
J114816.64+525150.3 (hereafter J1148+5251), further suggests a high star formation surface density 
of $\rm \sim1000\,M_{\odot}\,yr^{-1}\,kpc^{-2}$ over the central 
1.5 kpc region of the quasar host galaxy \citep{maiolino05,walter09}. 

Molecular CO (6-5) line emission has been detected in ten of the 
FIR luminous z$\sim$6 quasars (\citealp{bertoldi03b,walter03,carilli07,
wang10}, 2011, in prep.),   
indicating the existence of highly-excited molecular 
gas in the quasar hosts. The CO (3-2), (6-5), and (7-6) transitions 
detected in the z=6.42 quasar J1148+5251 reveal a molecular gas 
component on scales of $\sim$5 kpc in the host galaxy with CO 
excitation conditions similar to those found in local starburst 
galaxies and CO-detected quasars at lower redshifts \citep{bertoldi03b,walter04,riechers09}. 

Emission in the low-order CO transitions ($\rm J\leq2$) from the z$\sim$6 quasar host 
galaxies is poorly constrained due to the limited sensitivity and 
frequency coverage of the previous instruments \citep{wagg08,wang10}. 
The new Ka band receivers on the Expanded Very Large Array (EVLA, \citealp{perley11}) open an  
important frequency window for studies of the cold molecular gas in high-redshift 
galaxies (e.g., \citealp{ivison10,ivison11,riechers10}). In this paper, we report 
our EVLA observations of the CO (2-1) line emission in five z$\sim$6 
quasars \citep{fan04,fan06,willott10a,willott10b} 
Three of them are from the Sloan Digital Sky Survey (SDSS, \citealp{fan04,fan06}), 
with two objects, SDSS J084035.09+562419.9 and SDSS J092721.82+200123.7, 
previously detected in strong ($\rm >3\,mJy$) 250 
GHz dust continuum and molecular CO (6-5) and (5-4) line emission. 
Another object, SDSS J162331.81+311200.5, was detected in the [C {\small II}] 
line, but undetected in millimeter dust 
continuum and high-J CO transitions (Bertoldi et al. 2011, in prep.).
The other two objects are from the Canada-France High-z Quasar
Survey (CFHQS, \citealp{willott10a,willott10b}) and do not have published CO observations yet. 
One of them, CFHQS 142952.17+544717.6, was detected in 250 GHz dust continuum 
(Omont et al. 2011, in prep.). We describe the observations in Section 2, present 
the results in Section 3, and discuss the CO excitation and host galaxy 
evolution properties of the detections in Section 4. A $\rm \Lambda$-CDM
cosmology with $\rm H_{0}=71km\ s^{-1}\ Mpc^{-1}$, $\rm
\Omega_{M}=0.27$ and $\rm \Omega_{\Lambda}=0.73$ is adopted throughout this
paper \citep{spergel07}.

\section{Observations}

The observations were carried out using the Ka-band receiver on the EVLA 
in 2010 in the D, DnC, and C configurations. 
The WIDAR correlator in Open Shared Risk Observing mode provided a maximum 
bandwidth of 128 MHz and a resolution of 2 MHz in each of the two 
basebands (A/C and B/D intermediate frequency [IF] bands). The A/C IFs 
could not be tuned below 32 GHz. The redshifts and observing frequencies 
of the CO (2-1) line of the five targets are estimated with previous detections of the CO (6-5), 
[C {\small II}], or quasar UV lines (\citealt{carilli07,wang10};  
Bertoldi et al 2011, in prep.). For the 
three sources with redshifts of $\rm z\leq6.2$ [corresponding to redshifted CO (2-1) line 
frequencies of $\rm \nu_{obs}\geq32$ GHz], we use the two 128 MHz IF pairs overlapped 
by 30 MHz and cover a total bandwidth of 226 MHz (i.e., $\sim$2000 $\rm km\,s^{-1}$ 
in velocity and $\rm \sim0.05$ in redshift at $\rm z=6$)\footnote{10 MHz overlap and a total 
bandwidth of 246 MHz for J1429+5447.}. 
For the other two objects with $\rm z>6.2$, we 
centered the 128 MHz window of the B/D IF pairs on the line frequency 
and observed the continuum at $\geq$32 GHz with the other window. 
The observing time is 15 to 20 hours for each of the five targets (see Table 1).
Flux calibrations were performed using the standard VLA 
calibrators, 3C286 and 3C48, and we use 5-minute 
scan loops between targets and phase calibrators to calibrate the phase. 
The data were reduced with AIPS, and the  
spatial resolutions (FWHM) of the final images are typically 2$''$ for data taken in the 
D configuration and 0.7$''$ for the C configuration. 

\section{Result}

CO (2-1) line emission has been detected in two of the 
five z$\sim$6 quasars, J0927+2001 and J1429+5447, and marginally 
detected in J0840+5624. We present all the observing parameters and 
measurements in Table 1. The detailed results are listed below.

{\bf J0927+2001} Toward this source strong dust continuum 
at 850 GHz, 250 GHz, and 85 GHz, and CO (6-5) and (5-4) line emission 
were detected \citep{carilli07,wang10}. We have 
detected the CO (2-1) line and the emission distribution (averaged 
over a velocity range of 880 $\rm km\,s^{-1}$) along with a spectrum is 
shown in Figure 2. The line peak emission centeroid is consistent with the optical 
quasar position and the peaks of the high-J CO lines.  
The line width (FWHM) and redshift fitted with a single Gaussian 
profile are $\rm 590\pm130\,km\,s^{-1}$ and $\rm 5.7716\pm0.0012$
which are in good agreement with the measurements   
from the high-order CO transitions ($\rm z=5.7722\pm0.0006$ and 
$\rm FWHM=600\pm70\,km\,s^{-1}$, \citealt{carilli07}). 
The line emission appears marginally 
resolved by the $\rm 2.19''\times1.96''$ synthesized beam with a peak 
surface brightness of $\rm 147\pm21\, \mu Jy\,beam^{-1}$ and a total 
intensity of $\rm 230\pm45\,\mu Jy$, with a source size of 
$\rm (2.7''\pm0.4'')\times(2.4''\pm0.3'')$ determined from a fit with a two-dimensional 
Gaussian distribution (the deconvolved source size 
is about $\rm 1.7''\times 1.4''$, or $\rm 10\,kpc\times8\,kpc$). 
The corresponding line fluxes and luminosities (Table 1) are higher than the upper limits estimated from
previous GBT observations, but are still consistent given the large
uncertainties and baseline feature contamination in the GBT data \citep{wagg08,wang10}.

{\bf J1429+5447} Toward this object strong radio continuum 
emission was detected in the FIRST survey \citep{becker95} 
and recent VLBI observations \citep{frey11}, making it the
strongest radio source among the known z$\sim$6 quasars and the 
most distant radio-loud quasar. It has also been detected in 
dust continuum at 250 GHz with a flux density of $\rm \sim3$ mJy 
(Omont et al. 2011, in prep.). 
We have detected both CO (2-1) line emission and 
continuum emission at the line frequency. The 
continuum source is unresolved by the $\rm 0.71''\times0.67''$ synthesized 
beam and the flux density averaged over the line-free channels at 32 GHz 
is $\rm 257\pm15$ $\mu$Jy. We subtract the continuum by performing 
linear fitting to the visibility data, 
using the UVLIN task in AIPS. The CO line emission is resolved into two 
peaks with a spatial separation of $\sim$1.2$''$ (6.9 kpc at the
quasar redshift), and the optical and radio quasar positions are 
consistent with the west peak (Figure 3). 
A Gaussian fit to the spectra yields a redshift 
of $\rm z=6.1831\pm0.0007$ and a line width 
of $\rm FWHM=280\pm70\,km\,s^{-1}$ for the west source, 
and $\rm z=6.1837\pm0.0015$ and $\rm FWHM=400\pm140\,km\,s^{-1}$ for the east source. 
The line fluxes estimated with the peak surface brightness on 
the velocity-averaged map averaging over a velocity range of 
$\rm \sim450\,km\,s^{-1}$ are $\rm 0.065\pm0.011\,Jy\,km\,s^{-1}$ 
and $\rm 0.050\pm0.013\,Jy\,km\,s^{-1}$
for the west and east components, respectively. However, 
a two-dimensional Gaussian distribution fitted to the east component suggest 
possible extension with a source size of $\rm (1.1''\pm0.2'')\times(0.7''\pm0.2'')$, 
which should be checked with deeper observations at higher spatial resolution.

{\bf J0840+5624} This source was detected
in (sub)mm dust continuum emission and CO
(6-5) and (5-4) line emission; it has the broadest line width,
$\rm FWHM=860\,km\,s^{-1}$, among the CO-detected
z$\sim$6 quasars \citep{wang07,wang10}. We observed the line
at the redshift of $\rm z=5.8441\pm0.0013$ derived
from the high-order CO detections and find no clear detection
in a velocity-averaged map averaging over 1070 $\rm km\,s^{-1}$ made at the
full resolution of $\rm 1.09''\times0.76''$. At a lower resolution
of $\rm 2.19''\times1.96''$, marginal signal ($\rm 2.8\sigma$)
appears on the map (Figure 1), with a double-peaked
mophology along the east-west direction. The optical quasar position
is 0.8$''$ away from the east peak. We plot the spectrum at the position
of east peak in the right panel of Figure 1, and there is only very
marginal signal (1 to 2$\sigma$) over $\rm \sim-500$
to $\rm 500\,km\,s^{-1}$, i.e., the typical velocity range of the CO (6-5)
and (5-4) line emission \citep{wang10}.
The CO (2-1) line flux estimated with the surface
brightness of the east peak is $\rm 0.062\pm0.022\,Jy\,km\,s^{-1}$ (Table 1).
However, the signal is indeed marginal and deeper observations with a wider
bandwidth are required to improve the measurement.

{\bf J0210$-$0456} This object is the highest
redshift quasar known to date with $\rm z=6.438\pm0.004$ determined
from the object's $\rm Mg\,{\small II}\,\lambda2798\AA$ line emission \citep{willott10b}.
We searched for CO (2-1) line emission in the 128 MHz window centered at
the $\rm Mg\,{\small II}$ redshift but did not detect it.
Here we assume a line width of $\rm 800\,km\,s^{-1}$, which is the
typical full width at zero intensity ($\rm v_{FWZI}$) value found with
samples of high-z CO-detected quasars \citep{coppin08,wang10}
to estimate the upper limit of the line intensity.
The $\rm 1\sigma$ rms noise level on the map averaged over this velocity range
is $\rm \sigma_{rms} =16\,\mu Jy\,beam^{-1}$, and the $\rm 3\sigma$ upper limit of the line
flux is estimated as $\rm 3\sigma_{rms} v_{FWZI}=0.038\,Jy\,km\,s^{-1}$.
The corresponding 3$\sigma$ upper limit of
the line luminosity is $\rm {L'}_{CO(2-1)}<1.28\times10^{10}\,K\,km\,s^{-1}\,pc^2$ 
(see equation (3) in \citealt{solomon05}). However, we cannot rule out that 
the $\rm Mg\,{\small II}$ line emission is significantly offset 
from the quasar host galaxy redshift and CO (2-1) line falls outside the 128 MHz window. 
The continuum emission is 
also undetected with the other window centered at 32.1 GHz, and the 
channel-averaged map yields a 3$\sigma$ upper limit of $\rm <54\,\mu Jy$.

{\bf J1623+3112} This object is detected in $\rm [C\,{\small II}]$ 158$\mu$m  
fine structure line emission by Bertoldi et al. (2011, in prep.), 
but undetected in 250 GHz dust continuum \citep{wang07}. We searched for the CO (2-1) line in 
the 128 MHz-bandwidth window centered at the $\rm [C\,{\small II}]$ 
redshift of $\rm z=6.2605\pm0.0005$ and did not detect the line. 
The rms on the map averaged over a velocity range of 
$\rm 800\,km\,s^{-1}$ is $\rm 26\,\mu Jy\,beam^{-1}$. This yields 
a 3$\sigma$ upper limit of $\rm <0.062\,Jy\,km\,s^{-1}$ for the line flux  
and $\rm <2.0\times10^{10}\,K\,km\,s^{-1}\,pc^2$ for the line luminosity. 
The 3$\sigma$ upper limit of the continuum emission at 35 GHz 
measured with A/C IFs is $\rm <75\,\mu Jy$. 

\section{Discussion}

We have observed molecular CO (2-1) line emission toward five quasars 
at z$\sim$6 using the EVLA, and detections/marginal detection have been 
obtained from the three objects that have strong FIR dust 
continuum emission. This is consistent with the picture of massive star 
formation fueling by huge amount of molecular gas in these young 
quasar hosts. The detection of $\rm [C\,{\small II}]$ in J1623+3112 is 
also likely to be a sign of star formation, but the current sensitivity 
of our EVLA observations cannot detect molecular CO from the host galaxy. 
J0927+2001 and J0840+5624, were  
previously detected strongly in the CO (6-5) and (5-4) transitions. 
CO (2-1) line emission has been detected and marginally 
resolved in the host galaxy of J0927+2001 over a scale of $\sim$10 kpc. 
The molecular gas masses ($\rm M_{gas}$) estimated from the CO (2-1) line
peak surface brightness and the total intensity on the velocity-averaged
map are listed in Table 1, assuming a CO luminosity-to-gas mass conversion factor
of $\rm \alpha=0.8 M_{\odot}\,(K\,km\,s^{-1}\,pc^{2})^{-1}$ appropriate
for local ultraluminous infrared galaxies \citep{solomon97,downes98}.
These estimates are 1.7 and 2.5 times higher than the  
value of $\rm (1.8\pm0.3)\times10^{10}\,M_{\odot}$ estimated
from the high-order CO transitions \citep{carilli07,wang10}. 
We plot the CO excitation ladder of this source in Figure 4,
together with the results of Large Velocity Gradient (LVG) modeling
of the highly-excited molecular gas components 
(gas densities of 
order $\rm 10^{4}\,cm^{-3}$, kinetic temperatures of 50 to 60 K, 
and peak at $\rm J\geq6$) 
found in other high-z FIR and CO luminous
quasars and nearby starburst galaxies \citep{riechers06,riechers09,gusten06}.
We normalize the models to the high-order CO transitions. 
The CO (2-1) line flux measured with the peak surface 
brightness on the velocity-averaged map is consistent/marginally consistent 
with the values expected by these single-component models, while the total line flux 
integrated over the line-emitting area falls above all the models. This may suggest 
the exsitence of additional low excitation gas in the central $\rm \sim10\,kpc$ region 
as was found in the submillimeter galaxy AzTEC-3
at z=5.3 \citep{riechers10} and the nearby starburst 
galaxy M82 \citep{weiss05}. However, there are still large 
uncertainties in the measurements of all the three transitions, 
and observations of other CO transitions are necessary to 
address if there are multiple CO excitation components in the quasar host galaxy. 
Our observations show no evidence of excess CO (2-1) line emission and additional 
low excitation component in the host galaxy of J0840+5624. 

The C array imaging of the CO (2-1) line emission from J1429+5447 has 
resolved the molecular gas into two distinct peaks, with a spatial separation 
of $\sim$6.9 kpc; the quasar position is consistent with the West peak. 
There is no clear velocity offset ($\rm 26\pm60\,km\,s^{-1}$) between 
the two components. These results suggest a gas-rich, major merging 
system with two distinct components that are comparable in CO luminosity and 
molecular gas mass. The west component of this system 
is in a radio-loud quasar phase. Similar quasar-starburst systems with 
multiple CO emission peaks were previously found in 
the CO luminous quasars BRI 1202$-$0725 at z=4.7 \citep{omont96,carilli02}, 
BRI 1335$-$0417 at z=4.4 \citep{riechers08}  
and J1148+5251 at z=6.42 \citep{walter04}. These systems demonstrate  
the early phase of quasar-galaxy formation in which both AGN and starburst 
activities are triggered by major mergers and the molecular gas in the 
nuclear region is not fully coalesced \citep{narayanan08}. We will 
expect further high-resolution observations with the EVLA in C or B 
array to constrain the gas surface density and dynamics, 
and with ALMA or the PdBI to resolve the dust continuum and distributed 
star formation in these young quasar host galaxies.

\acknowledgments 
This work is based on observations carried out with the Expanded Very Large 
Array (NRAO). The National
Radio Astronomy Observatory (NRAO) is a facility of the National
Science Foundation operated under cooperative agreement by Associated
Universities, Inc. We acknowledge support from the Max-Planck Society
and the Alexander von Humboldt Foundation through the Max-Planck-Forschungspreis
2005. Dominik A. Riechers acknowledges support from NASA through Hubble
Fellowship grant HST-HF-51235.01 awarded by the Space Telescope Science
Institute, which is operated by the Association of Universities for
Research in Astronomy, Inc., for NASA, under contract NAS 5-26555.
M. A. Strauss acknowledges the support of NSF grant Ast-0707266.
{\it Facilities:} \facility{EVLA}

\begin{table}
{\scriptsize \caption{Observations}
\begin{tabular}{lccccccc}
\hline \noalign{\smallskip} 
Name & $\rm t_{obs}$ & Configuration & Redshift & FWHM & I$\Delta$v & $\rm {L'}_{CO(2-1)}$ & $\rm M_{gas}$ \\
     & hour &    &  & $\rm km\,s^{-1}$ & $\rm Jy\,km\,s^{-1}$ & $\rm 10^{10}\,K\,km\,s^{-1}\,pc^{2}$ & $\rm 10^{10}\,M_{\odot}$\\ 
 (1) & (2) & (3) & (4) & (5) & (6) & (7) & (8) \\
\noalign{\smallskip} \hline \noalign{\smallskip}
\multirow{2}{*}{SDSS J092721.82+200123.7} & \multirow{2}{*}{20} & \multirow{2}{*}{D, DnC} & \multirow{2}{*}{5.7716$\pm$0.0012} & \multirow{2}{*}{590$\pm$130} & $\rm 0.129\pm0.018^{c}$&3.70$\pm$0.52$^{c}$ & 3.0$\pm$0.4$^{c}$ \\
     &       & &     &  &   0.202$\pm$0.040$^{d}$ & 5.80$\pm$1.15$^{d}$ & 4.6$\pm$0.9$^{d}$\\
CFHQS J142952.17+544717.6W$^{e}$ & 15 & C &6.1831$\pm$0.0007 & 280$\pm$70 & 0.065$\pm$0.011 &2.06$\pm$0.35 &1.6$\pm$0.3 \\
CFHQS J142952.17+544717.6E & 15 & C &6.1837$\pm$0.0015 & 400$\pm$140 & 0.050$\pm$0.013&1.59$\pm$0.41 & 1.3$\pm$0.3\\
SDSS J084035.09+562419.9 & 20 & D, DnC, C & 5.8441$^{b}$ & 860$^{b}$ & $\rm 0.062\pm0.022$ & 1.81$\pm$0.64 & 1.4$\pm$0.5 \\
CFHQS J021013.19$-$045620.9 & 15& C & 6.438$^{a}$ & -- & $<$0.038 & $<$1.28 & $<$1.0\\
SDSS J162331.81+311200.5 & 20 & D & 6.26$^{f}$ &  -- & $\rm <0.062$ & $\rm <2.00$ & $<$1.6\\
\noalign{\smallskip} \hline
\end{tabular}\\
$^{a}$Redshift measured from the quasar $\rm Mg\,{\small II}$ line 
emission (Willott et al. 2010); $^{b}$Redshift and CO line width measured 
from the CO (6-5) and (5-4) lines (Wang et al. 2010). $^{c}$CO (2-1) line flux 
and luminosity derived from the peak surface brightness on the 
velocity-averaged map; $^{d}$CO (2-1) line flux and luminosity 
derived from the total intensity integrated over the 
line-emitting area on the velocity-averaged map. $^{e}$The West 
component of J1429+5447 is consistent with the optical 
quasar position. $^{f}$Redshift from the $\rm C\,{\small II}$ 
line detection (Bertoldi et al. 2011 in prep.)
}
\end{table}
\begin{figure}
\plottwo{fig1a.eps}{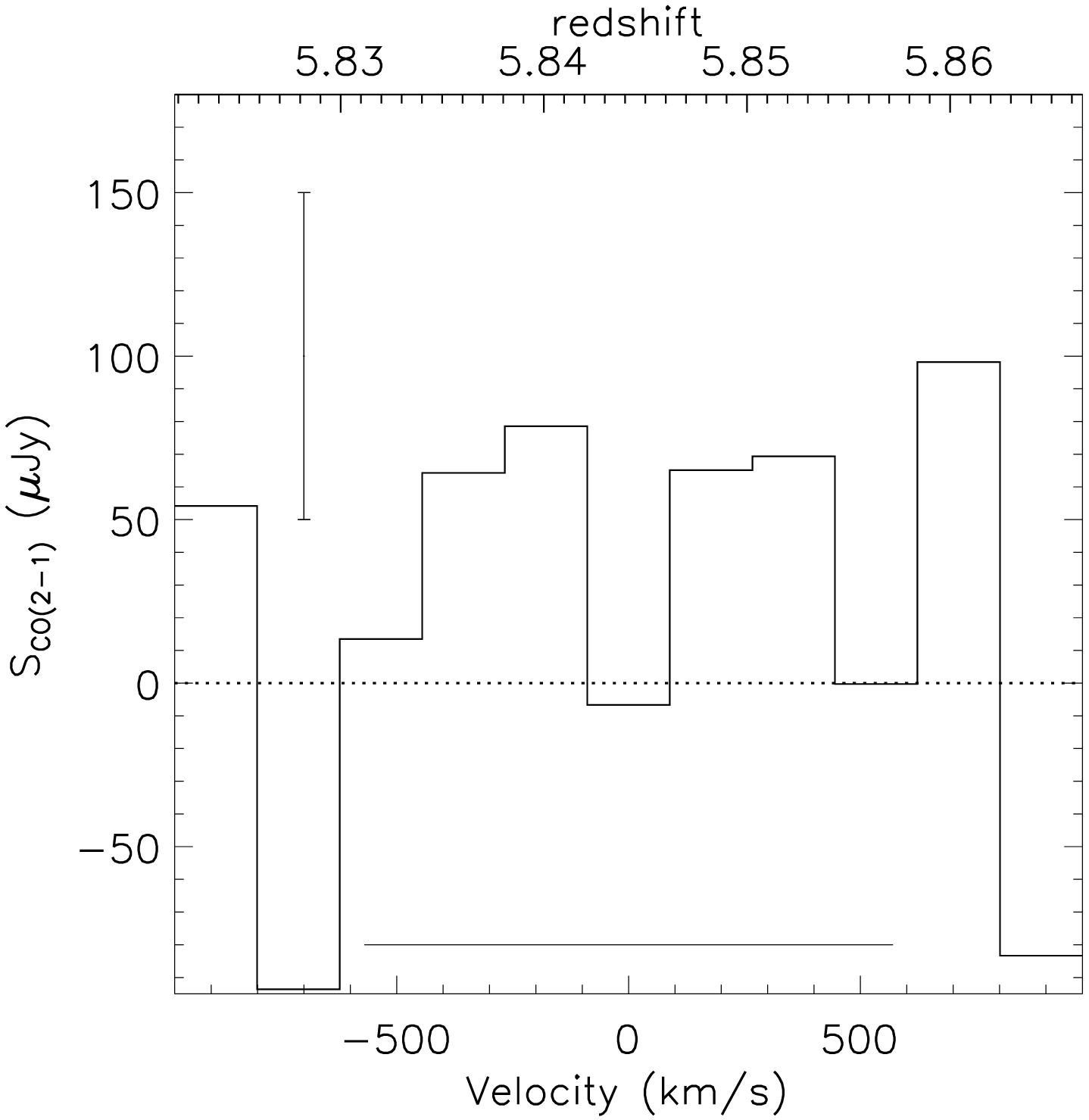}
\caption{{\it Left}--Velocity-averaged map of the CO (2-1) line 
emission from J0840+5624 at a resolution of $\rm 2.19''\times1.96''$, 
averaged over $-$535 to 535 $\rm km\,s^{-1}$. 
The contour levels are (-2, 2, 3)$\times$18 $\rm \mu Jy\,beam^{-1}$, 
and the 1$\rm \sigma$ on the map is 21$\rm \mu Jy\,beam^{-1}$.
The cross denotes the optical position of the quasar. {\it Right}--Spectrum 
at the position of the east peak, binned to a channel width of 
20 MHz ($\rm 178\,km\,s^{-1}$). The zero 
velocity corresponds to a redshift of $\rm z=5.8441$ measured from the 
high-order CO emission (Wang et al. 2010). 
The error bar denotes the typical $\rm \pm1\sigma$ rms value. 
The dotted line shows the zero intensity level and the solid line at the 
bottom denotes $\rm v_{FWZI}$ of 
the high-J CO lines (Wang et al. 2010).}
\end{figure}
\begin{figure}
\plottwo{fig2a.eps}{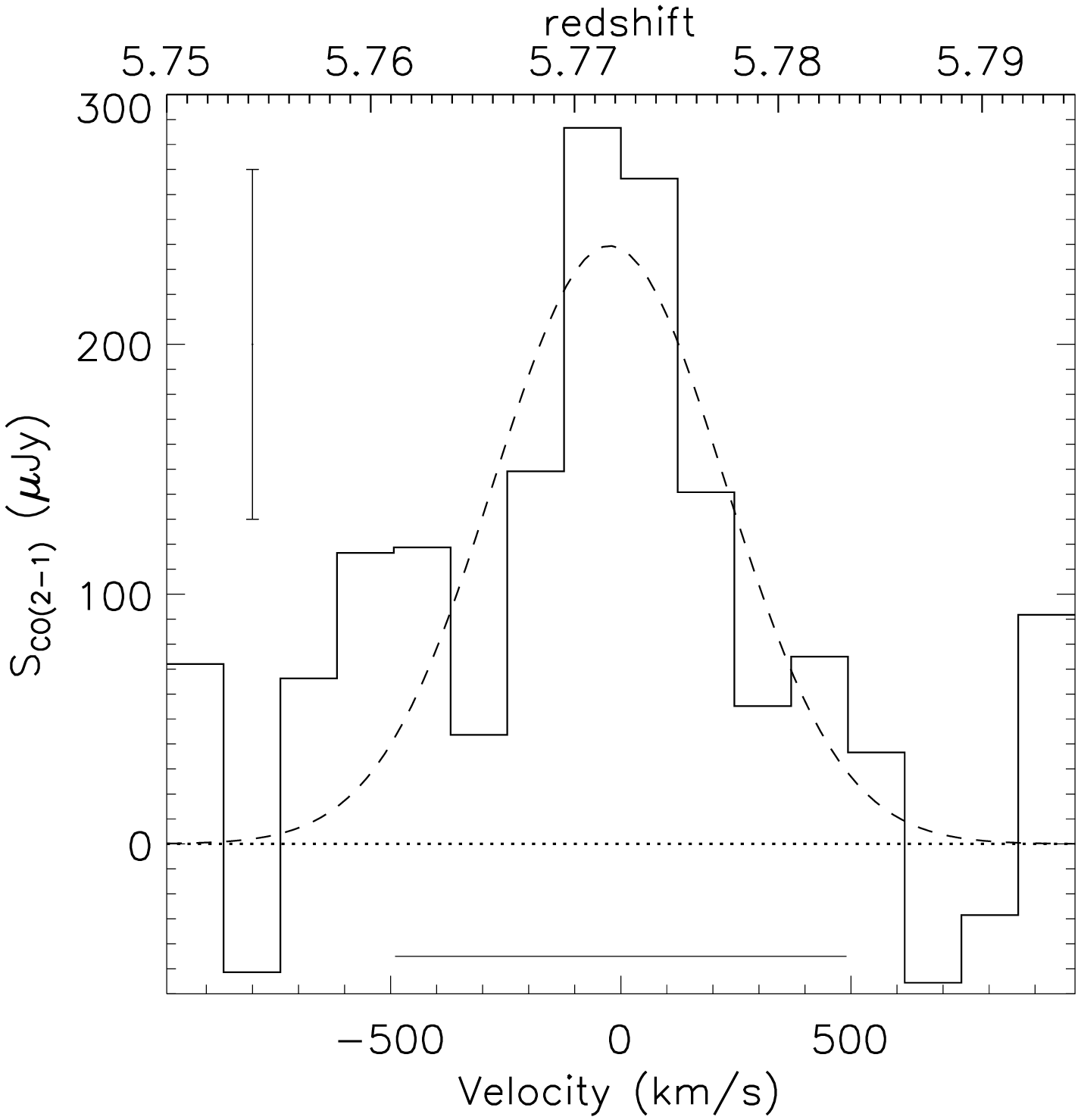}
\caption{{\it Left}--Velocity-averaged map of the CO (2-1) line
emission from J0927+2001 at a resolution of $\rm 2.19''\times1.96''$, 
averaged over $\sim-$440 to 440 $\rm km\,s^{-1}$.
The contour levels are (-2, 2, 3, 4, 5)$\times$21 $\rm \mu Jy\,beam^{-1}$.
The cross denotes the optical quasar position. 
The source appears marginally resolved with a source size of $\rm 2.7''\times 2.4''$ 
fit to a two-Dimensional Gaussian profile (deconvolved size of $\rm 1.7''\times 1.4''$, 
or $\rm 10\,kpc\times8\,kpc$). {\it Right}--Spectrum integrated over $\sim$2 beam area. 
The spectrum is binned to a channel width of 14 MHz ($\rm 123\,km\,s^{-1}$) and the zero
velocity corresponds to the high-J CO redshift 
of $\rm z=5.7722$ (Carilli et al. 2007). The error bar represents  
the typical $\rm \pm1\sigma$ rms in each channel. 
The dashed line denotes a 
Gaussian fit to the line and the dotted line shows the zero 
intensity level. The solid line at the 
bottom shows $\rm v_{FWZI}$ of the high-J CO lines (Carilli et al. 2007).}
\end{figure}
\begin{figure}
\includegraphics[height=2.0in]{fig3a.eps}
\vskip -2.0in
\hspace*{2.2in}
\includegraphics[height=2.0in]{fig3b.eps}
\vskip -2.0in
\hspace*{4.4in}
\includegraphics[height=2.0in]{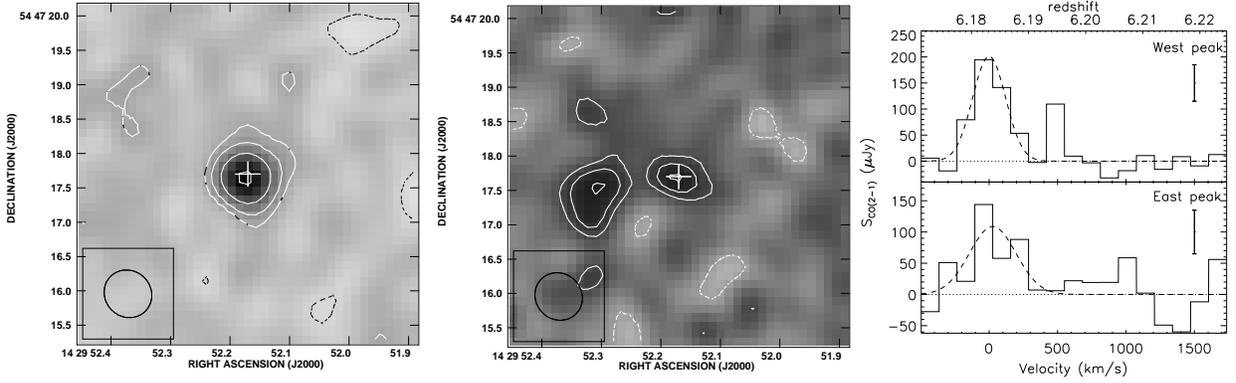}
\caption{{\it Left}--32 GHz continuum emission from J1429+5447 at a 
resolution of $\rm 0.71''\times0.67''$. 
The contour levels are (-2,2,4,8,16)$\times$15 $\rm \mu Jy\,beam^{-1}$. 
The cross denotes the optical position of the quasar.
Middle--Velocity-averaged map of the CO (2-1) line
emission, averaged from $\sim-$100 to 340 $\rm km\,s^{-1}$ to get best S/N for both components.
The contour levels are (-2, 2, 3, 4)$\times$27 $\rm \mu Jy\,beam^{-1}$.
The source is resolved into two peaks with a separation of $\sim$1.2$''$ (6.9 kpc), 
and the quasar position is consistent with the west peak.
{\it Right}--Spectra of the two components on the map. 
The quasar continuum has been subtracted from the West component.
We adopt the surface brightnesses at the peak positions as the
measurements of the line flux densities in each channel. 
The spectra are binned to a channel width of 14 MHz ($\rm 130\,km\,s^{-1}$) and the zero
velocity corresponds to the redshift of $\rm z=6.1831$ measured with 
the west component. The dashed lines denote a
Gaussian fit and the dotted lines show the 
zero intensity level. The spike at the 8th channel (i.e., $\rm \sim500\,km\,s^{-1}$) of 
the upper panel is due to the poor sensitivity ($\rm 1\sigma$ rms 
of $\sim160\,\mu Jy\,beam^{-1}$) at the edge 
of the A/C IFs. The error bar represents
the typical $\rm \pm1\sigma$ rms in the rest channels.}
\end{figure}
\begin{figure}
\plotone{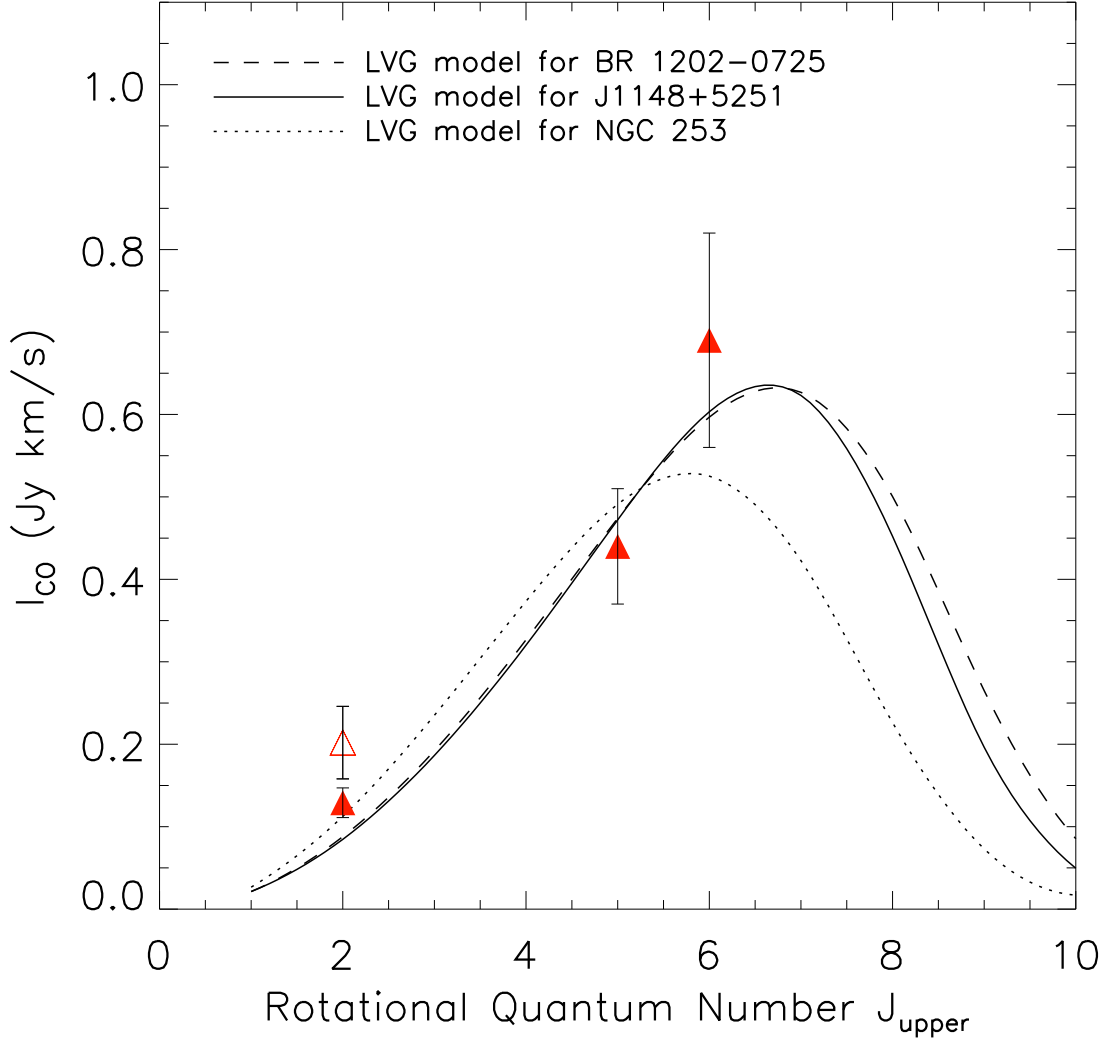}
\caption{CO excitation ladder of J0927+2001. The filled 
triangle at J=2 is the CO (2-1) line flux measured with the peak surface brightness on 
the velocity-averaged map, while the open triangle is derived with the total 
intensity integrated over the line-emitting area of $\rm 2.7''\times 2.4''$.  
The error bars show the 1$\sigma$ uncertainties of the line fluxes.
The solid and dashed lines show the LVG models of J1148+5251 
with kinetic temperature, $\rm T_{kin}=50\,K$, molecular hydrogen density, 
$\rm \rho_{gas}(H_2)=10^{4.2}\,cm^{-3}$ from Riechers et al. (2009)
and BR 1202-0725 with $\rm T_{kin}=60\,K$, $\rm \rho_{gas}(H_2)=10^{4.1}\,cm^{-3}$ 
from Riechers et al. (2006), respectively. The dotted line represents the LVG model 
of the central region of the nearby starburst galaxy NGC 253 
with $\rm T_{kin}=60\,K$, $\rm \rho_{gas}(H_2)=10^{3.9}\,cm^{-3}$ 
from G$\rm \ddot{u}$sten et al. (2006). These models are fitted to 
the CO (6-5) an (5-4) transitions. We have not made any correction 
for the continuum emission at the CO (2-1) line frequency as there 
is no significant detection. The continuum contimination to 
the CO (2-1) line flux is estimated to be $\leq$5\% based on the FIR dust 
continuum determined with previous (sub)mm detections (Wang et al. 2010).
}
\end{figure}

\end{document}